\documentstyle[12pt]{article}
\begin{document}
\vspace{1.cm}
\begin{center}
\    \par
\    \par
\    \par
\     \par

              {\bf{THE LIGHT FRONT ANALYSIS OF $\pi^{-}$ MESONS PRODUCED
           IN THE RELATIVISTIC NUCLEUS-NUCLEUS COLLISIONS}}
\end{center}
\   \par
\   \par
\   \par
\   \par
\    \par
\par
    {\bf{ M.Anikina$^{1}$,    L.Chkhaidze$^{2}$,    T.Djobava$^{2}$,}}\par
    {\bf{ V.Garsevanishvili $^{3,4}$,     L.Kharkhelauri$^{2}$}}\par
\  \par
$^{1}$ Joint Institute for Nuclear Research, 141980 Dubna, Russia\par
$^{2}$ High Energy Physics Institute, Tbilisi State University,\par
   University Str. 9, 380086 Tbilisi, Republic of Georgia\par
$^{3}$ Mathematical Institute of the Georgian Academy of Sciences \par
   M.Alexidze Str. 1 , 380093 Tbilisi, Republic of Georgia  \par
$^{4}$ CERN, CH-1211, Geneva 23, Switzerland\par
\   \par
\   \par
E-mail: djobava@sun20.hepi.edu.ge     \par
\pagebreak
\   \par
\   \par
\begin{center}

                     \bf{ ABSTRACT }
\end{center}
\ \par
\par
The light front analysis of $\pi^{-}$ mesons in
$A_{P}$ (He, C, Mg, O) + $A_{T}$
(Li,C, Ne, Mg,Cu,Pb) collisions is carried out. The phase space of secondary
pions is naturally divided into two parts in one of which the thermal
equilibration assumption seems to be in a good agreement with data. Corresponding
temperatures are extracted and compared to the results of other
experiments.
The dependence of the average temperature $T$
on $(A_{P}*A_{T})^{1/2}$ is studied.
    \par
\    \par
\    \par
\    \par
\begin{sloppypar}

\    \par
\    \par
%\tabcolsep=6.mm
\begin{tabular}{|l|}\hline
\    \\
{\bf{NUCLEAR REACTION}}
The light front analysis of $\pi^{-}$ mesons in \\
$A_{P}$ (He, C, Mg, O) + $A_{T}$
(Li,C, Ne, Mg,Cu,Pb) collisions is carried out.\\
In some region of phase space of $\pi^{-}$
mesons the thermal equilibration seems \\to be reached,
 which is characterized by the temperature $T$.
\\
\     \\
\hline
\end{tabular}
\end{sloppypar}
\pagebreak
\begin{center}
\bf { 1.  INTRODUCTION }
\end{center}
\  \par
\par
In the experiments with beams of relativistic heavy ions one hopes
to observe the extreme conditions when the phase transitions in nuclear
matter are expected (see, e.g. [1-3]).
\par
   For the experimental study of such  transitions it is necessary to
understand the  mechanism  of  collisions  and investigate
the characteristics of multiparticle production  in nucleus-nucleus
interactions. The study of single particle inclusive processes is
one of the simplest and effective tools for the understanding of
dynamics of multiple production of secondaries.
\par
   In this  respect  it  is important  to  investigate  the
properties of $\pi^{-}$ mesons, which are predominantly produced particles
carrying the information about the dynamics of collision and which are
reliably  identified.  Besides,  the  pion   production   is   the
predominant production process at Dubna energies.
\par
   Our previous results on pion production experiment (cross-sections,
multiplicities, rapidities, transverse momenta, intercorrelations
between various characteristics, etc) using  the
streamer chamber spectrometer  SKM-200 and its modified version GIBS
   in  inelastic  and  central
nucleus-nucleus interactions are presented in [4-6].
\par
In this paper we present the light front analysis of $\pi^{-}$ mesons
produced in He-Li, He-C, C-Ne, Mg-Mg, C-Cu and O-Pb collisions. In some cases
this type of analysis [7] seems to be more sensitive to the details of
the interaction mechanism as compared to the presentation of data in
terms of the well known variables $x_{F}$, rapidity etc.
\  \par
\  \par
\begin{center}
\bf { 2.  EXPERIMENT }
\end{center}
\  \par
\par
    The data were obtained using the  SKM-200  facility
and its modified version GIBS of the Dubna
Joint Institute for Nuclear Research.  SKM-200-GIBS   consists  of  a   2m
streamer chamber, placed in a magnetic field of  $\sim$ 0.8  T  and   a
triggering system. The streamer chamber was exposed by beams  of
He , C , O ,  Ne  and  Mg  nuclei accelerated in the synchrophasotron  up
to a momentum of  4.5 GeV/c  per incident nucleon . The solid targets
in the form of thin discs with thickness  0.2$\div$0.4 g/cm$^{2}$
( for  Li the thickness was  1.59 g/cm$^{2}$ and for Mg 1.56 g/cm$^{2}$)
were mounted within the fiducial
volume
of the chamber. Neon gas filling of the chamber served also  as  a
nuclear target. The triggering system  allowed  the  selection  of
"inelastic"  and "central" collisions.
\par
   The inelastic trigger selected all inelastic  interactions of
the incident nuclei on the target.
%\par
   The central trigger selected events with no charged projectile
spectator fragments (with  $P/Z>3$ GeV/c ) within a cone of  half angle
$\Theta_{ch}$ = 2.4$^{0}$ or  2.9$^{0}$   (the trigger efficiency was
99$\%$ for events  with  a single charged particle in the cone).
Later a neutron detector was added  to  the  veto system for excluding
events with spectator neutrons in a cone  of half angle $\Theta_{n}$ =
1.8$^{0}$ or  2.8$^{0}$(the  trigger
efficiency  was   80$\%$  for events with a single neutron in the cone).
The  trigger  mode  for each exposure is defined as  T ($\Theta_{ch}$
,$\Theta_{n}$ ) ($\Theta_{ch}$  and  $\Theta_{n}$  expressed  in degrees and rounded
to the closest  integer  value).  Thus   T(0,0)  corresponds to  all
inelastic  collisions.  For   He-Li   and He-C  collisions we had
only an inelastic trigger.
The experimental setup and the logic of the triggering systems are
presented in Fig.1.
\par
  Primary results of scanning and measurements were biased due to several
experimental effects and appropriate corrections were introduced. The biases
and correction procedures were discussed in detail in [4,5].
  Average measurement errors of the  momentum  and
production angle  determination for $\pi^{-}$   mesons are $<\Delta P/P
>$= 5$\%$, $\Delta$$\Theta$ =0.5$^{0}$ for He-Li, He-C, C-Ne, C-Cu, O-Pb and
$<\Delta P/P
>$= 1.5$\%$, $\Delta$$\Theta$ =0.1$^{0}$ for Mg-Mg.
%\newpage
\  \par
\  \par
\  \par
\begin{center}

     \bf{ 3.  LIGHT FRONT PRESENTATION OF INCLUSIVE DISTRIBUTIONS}
\end{center}
\   \par
\par
The study of single particle inclusive processes [8] remains one of the
simplest and effective tools for the investigation of multiple production of
secondaries at high energies.
An important role in establishing of many properties of multiple production
is played by the choice of kinematical variables in terms of which observable
quantities are presented (see in this connection, e.g. [9]). The
variables which are commonly used are the following:
%\hfil\break
 the Feynman
$x^{}_F=2p^{}_z/\sqrt{s}$,
 rapidity $y= {1\over{2}}{\rm ln}
[(E+p^{}_z)/(E-p^{}_z)]$,
transverse scaling variable $x^{}_T=2p^{}_T/\sqrt{s}$ etc.
In the case of azimuthal symmetry the surfaces of const $x^{}_F$ are the
planes $p^{}_z=x^{}_F\sqrt{s}/2$, surfaces of constant $y$ are the
hyperboloids
\begin{eqnarray*}
p^2_z\left[\left({1+e^{2y}\over{1-e^{2y}}}\right)^2-1\right]-p^2_T=m^2
\end{eqnarray*}

\noindent and the surfaces of constant $x^{}_T$ are the straight lines
$p^{}_T=x^{}_T\sqrt{s}/2$ in the phase space.

Here we propose  unified scale invariant variables for the presentation of single
particle inclusive distributions, the properties of which are described
below.

Consider an arbitrary 4--momentum $p^{}_{\mu}(p^{}_0,\vec p)$ and introduce the light
front combinations [10]:
\begin{eqnarray}
p^{}_{\pm}=p^{}_0\pm p^{}_3 \label{eq1}
\end{eqnarray}

If the 4--momentum $p^{}_{\mu}$ is on the mass shell $(p^2=m^2)$, the combinations
$p^{}_{\pm},\ \vec p^{}_T$ (where $\vec p^{}_T=(p^{}_1,\ p^{}_2)$) define the so called
horospherical coordinate system (see, e.g. [11]) on the
corresponding mass shell hyperboloid
$p^2_0-\vec p\, ^2=m^2$.

Let us construct the scale invariant variables [7]:
\begin{eqnarray}
\xi^{\pm}=\pm {p^c_{\pm}\over{p^a_{\pm}+p^b_{\pm}}} \label{eq2}
\end{eqnarray}

\noindent in terms of the 4--momenta $p^a_{\mu},\ p^b_{\mu},\ p^c_{\mu}$ of particles
$a,\ b,\ c$, entering the inclusive reaction $a+b\to c+X$.
The $z$-axis is taken to be the collision axis, i.e. $p_{z}=p_{3}=p_{L}$.
 Particles $a$
and $b$ can be hadrons, heavy ions, leptons.
The light front variables $\xi^{\pm}$ in the centre of mass frame
are defined as follows [7]:
\begin{eqnarray}
\xi^{\pm}&=&\pm {E\pm p_z\over{\sqrt{s}}}=\pm {E+|p_z|\over{
\sqrt{s}}}  \label{eq3}
\end{eqnarray}
where $s$ is the usual Mandelstam variable,
$E=\sqrt{p^{2}_z+p^{2}_T+m^{2}}$ and $p_{z}$ are
the energy and the $z$ - component of the momentum of produced particle.
The upper sign in Eq. (3) is used for the right hand side hemisphere and
the lower sign for the left hand side hemisphere. It is convenient also to
introduce the variables
\begin{eqnarray}
\zeta^{\pm}=\mp{\rm ln}|\xi^{\pm}| \nonumber
\end{eqnarray}
in order to enlarge the scale in the region of small $\xi^{\pm}$.

The invariant differential cross section in terms of these variables
looks as follows:
\begin{eqnarray}
E{d\sigma\over{d\vec p}}={|\xi^{\pm}|\over{\pi}}\ {d\sigma\over{
d\xi^{\pm}dp^{2}_T}} = {1\over{\pi}}\ {d\sigma\over{
d\zeta^{\pm}dp^{2}_T}}  \label{eq4}
\end{eqnarray}
\par
Consider two limiting cases:\\
%\itemitem{1)}
1) $|p_z|\gg p_T$ - fragmentation region, according to
the common terminology.\\
%\itemitem{}
In this case:
\begin{eqnarray}
\xi^{\pm}\rightarrow {2p_z\over{\sqrt{s}}}=x^{}_F
\label{eq5}
\end{eqnarray}
%\itemitem{2)}
2) $p_T\gg |p_z|$ - high $p^{}_T$-region.\\
%\itemitem{}
In this case:
\begin{eqnarray}
\xi^{\pm}\rightarrow {m_T\over{\sqrt{s}}}\rightarrow
{p_T\over{\sqrt{s}}}={x^{}_T\over{2}}\ ;\ m_T=\sqrt{p_T^{2}+m^{2}}
\label{eq6}
\end{eqnarray}

Thus, in these two limiting regions of $\xi^{\pm}$--variables
go over to the well known variables $x^{}_F$ and $x^{}_T$, which are
intensively used in high energy physics.

$\xi^{\pm}$--variables are related to $x^{}_F$,\ $x^{}_T$ and rapidity
$y$ as follows:
\begin{eqnarray}
\xi^{\pm}&=&{1\over{2}}\left(x^{}_F\pm \sqrt{x^2_F+x^2_{T}}\right)\
;\ x^{}_{T}={2m_T\over{\sqrt{s}}} \label{eq7}\\
y&=&\pm {1\over{2}}\ {\rm ln}\, {(\xi^{\pm}\sqrt{s})^2\over{m^{2}_T}}
%~;~ m_{T}=\sqrt{p^{2}_{T}+m^{2}}
\label{eq8}
\end{eqnarray}
\par
The principal differences of $\xi^{\pm}$ distributions as compared to the
corresponding $x_{F}$ -- distributions are the following:
1) existence of some forbidden region around the point $\xi^{\pm}=0$,
2) existence of maxima at some $\tilde{\xi^{\pm}}$ in the region of
 relatively small $|\xi^{\pm}|$, 3) existence of the limits for
$\vert\xi^{\pm}\vert \leq m/\sqrt{s}$.
 The maximum
at $\tilde{\zeta}^{\pm}$ is also observed in the invariant differential cross
section
$\displaystyle {1\over{\pi}}\ {d\sigma\over{d\zeta^{\pm}}}$. However, the region
$|\xi^{\pm}|>|\tilde{\xi}^{\pm}|$ goes over to the region
$|\zeta^{\pm}|<|\tilde{\zeta}^{\pm}|$ and vice
versa.
\par
Note that the light front variables have been introduced long time ago by
Dirac [10] and they are widely used now in the treatment of
many theoretical problems
(see, e.g. original and review papers [11-20] and references therein). They have
been used also in a number of phenomenological applications (see, e.g. [21]).

%\newpage
\  \par
\  \par
\  \par
\begin{center}
\bf{4.  THE ANALYSIS OF PION DISTRIBUTIONS IN TERMS OF LIGHT FRONT
VARIABLES}
\end{center}
\  \par
\  \par
\par
The analysis has been carried out for the $\pi^{-}$ mesons from
He(Li,C), C-Ne, Mg-Mg, C-Cu and O-Pb collisions.
In Figs.2,3 the $x_{F}$ -- and $\xi^{\pm}$ -- distributions of $\pi^{-}$ mesons
from He(Li,C), C-Cu and O-Pb interactions
 are presented.  These distributions are similar for all analysed
pairs of nuclei.
The experimental data
for invariant distributions $(1/\pi) \cdot dN/d\zeta^{\pm}$ in He(Li,C), C-Cu
and O-Pb collisions are shown in Fig.4. The curves are the result of
the polynomial
approximation of the experimental distributions and the
maxima are observed in these distributions
 at $\zeta^{\pm}=\tilde{\zeta^{\pm}}$ ($\tilde{\zeta^{\pm}}=2.0\pm0.3$
for He(Li,C), Mg-Mg,  C-Cu and O-Pb; $\tilde{\zeta^{\pm}} =2.1\pm0.3$
for C-Ne). The values of $\tilde{\zeta^{\pm}}$
are the boundaries of the two regions with significantly
different characteristics of
secondaries.
\par
In Figs.5(a,b,c), 6(a,b,c) the $p_{T}^{2}$ and the angular
 distributions of $\pi^{-}$ mesons
from  He(Li,C),
Mg-Mg, C-Cu and O-Pb interactions in different regions of $\zeta^{+}$ ( $\zeta^{+} >
\tilde{\zeta^{+}}$
and $\zeta^{+} < \tilde{\zeta^{+}}$) in the forward hemisphere are presented.
Similar results have been
obtained for the backward emitting $\pi^{-}$ mesons and
for all analysed pairs of nuclei.
\par
One can see from Figs.5,6, that the
$p_{T}^{2}$ and the angular distributions of $\pi^{-}$ mesons differ significantly in
$\zeta^{+} > \tilde{\zeta^{+}}$  and $\zeta^{+} < \tilde{\zeta^{+}}$
regions. The angular distribution of pions in the region $\zeta^{+} <
\tilde{\zeta^{+}}$ (Figs.6.a and 6.c)
is sharply anisotropic in contrast to the almost flat distribution
in the region $\zeta^{+} > \tilde{\zeta^{+}}$ (Figs.6.a and 6.b).
The slopes of
$p_{T}^{2}$ -- distributions differ greatly in different regions of
$\zeta^{\pm}$ (Fig.5.a, 5.b, 5.c).
The average values $<p_{T}^{2}>$ in these two regions also differ
(e.g. for $\pi^{-}$ mesons from Mg-Mg:
$<p_{T}^{2}>=(0.027\pm0.002$) (GeV/c)$^2$ in the region  $\zeta^{+} > \tilde{\zeta^{+}}$;
$<p_{T}^{2}>=(0.103\pm0.009$)  (GeV/c)$^2$ in
the region $\zeta^{+} < \tilde{\zeta^{+}}$).
\par
The flat behaviour of the angular distribution allows one to think that
one observes a partial thermal equilibration in the region
 $|\zeta^{\pm}| > |\tilde{\zeta^{\pm}}|$ ($|\xi^{\pm}| < |\tilde{\xi^{\pm}}|$)
of phase space.
\par
Note, that the surfaces of constant $\xi^+$ are the paraboloids
\begin{eqnarray}
p_z={p^{2}_T+m^{2}-(\xi^{+}\sqrt{s})^2\over{-2\xi^{+}\sqrt{s}}}
\label{eq9}
\end{eqnarray}
\noindent in the phase space.
Thus the paraboloid
\begin{eqnarray}
p_z={p^{2}_T+m^{2}-(\tilde{\xi}^{+}\sqrt{s})^2\over{-2\tilde{\xi}^{+}
\sqrt{s}}} \label{eq10}
\end{eqnarray}

\noindent separates two groups of
particles with significantly different characteristics.
\par
To describe the spectra in the region $\zeta^{+} > \tilde{\zeta^{+}}$ the
Boltzmann
\begin{center}
$  f(E)\sim e^{-E/T} $
\end{center}
and the Bose-Einstein (B-E)
\begin{center}
$  f(E)\sim (e^{-E/T} - 1)^{-1}$
\end{center}
distributions have been used.
\par
 The distributions
$1/\pi \cdot  dN/d\zeta^{+}$, $dN/dp_{T}^{2}$, $dN/dcos\Theta$ look in
this region as follows~:
\begin{eqnarray}
{1\over{\pi}}\ {dN\over{d\zeta^+}}&\sim&\int_0^{p^2_{T,max}}Ef(E)dp^2_T
\label{eq11}\\
{dN\over{dp^2_T}}&\sim&\int_0^{p^{}_{z,max}}f(E)dp^{}_z \label{eq12}\\
{dN\over{d\cos\theta}}&\sim&\int_0^{p^{}_{max}}f(E)p^2dp \label{eq13}\\
E&=&\sqrt{\vec p\, ^2+m^2_{\pi}}\ ,\ \vec p\, ^2=p^2_z+p^2_T \label{eq14}
\end{eqnarray}
where:
\begin{center}
${p^{2}_{T,max}} =(\tilde{\xi^{+}}\sqrt{s})^{2} - m_{\pi}^{2}$
\end{center}
\begin{center}
${p_{z,max}} =[p_{T}^{2}+m^{2}-(\tilde{\xi^{+}}\sqrt{s})^{2}]/
(-2\tilde{\xi^{+}}\sqrt{s})$
\end{center}
\begin{center}
$p_{max}=(-\tilde{\xi^{+}}\sqrt{s}cos\Theta + \sqrt{(
\tilde{\xi^{+}}\sqrt{s})^{2}- m_{\pi}^{2}
sin^{2}\Theta})/sin^{2}\Theta$
\end{center}
\par
The experimental distributions in the region $\zeta^{+} >\tilde{\zeta^{+}}$
have been fitted by the expressions (11), (12), (13), respectively. The results
of the fit are given in Table 1 and Figs. 5-7.
They show a rather good agreement with experiment.
\par
In the region $\zeta^{+} < \tilde{\zeta^{+}}$ the $p_{T}^{2}$ -- distributions
has been fitted by the formula
\begin{eqnarray}
\frac {dN}{dp_{T}^{2}} \sim \alpha \cdot e^{-\beta_{1}P_{T}^{2}} +
(1-\alpha) \cdot e^{-\beta_{2}p_{T}^{2}} \label{eq15}
\end{eqnarray}
and the $\zeta^{+}$ -- distributions by the formula
\begin{eqnarray}
\frac{1}{\pi}\cdot\frac {dN}{d\zeta^{+}} \sim (1 - \xi^{+})^{n}=
(1 - e^{-\vert \zeta^{+}\vert})^{n} \label{eq16}
\end{eqnarray}
%\par
 The results of the fit are given in
Table 2 and Figs.5 and 7.
\par
Thus the spectra of $\pi^{-}$ mesons in the region $\zeta^{+} >
\tilde{\zeta^{+}}$ are
satisfactorily described by the formulae which follow from
the thermal equilibration. The same formulae when extrapolated to the region
$\zeta^{+} < \tilde{\zeta^{+}}$ (Fig.7.a for Mg-Mg and 7.b for C-Cu)
deviate significantly from the data.
On the other hand the dependence $(1 - \xi^{+})^{n}$ is in a good
agreement with experiment in the region $\zeta^{+} < \tilde{\zeta^{+}}$ and
deviates from it in the region $\zeta^{+} > \tilde{\zeta^{+}}$ (Fig.7.a and
7.b).
\par
In Fig.8 the dependence of the average temperature $T$ (averaged over the
values of $T$ obtained from the fitting the distributions
$1/\pi \cdot  dN/d\zeta^{+}$, $dN/dp_{T}^{2}$, $dN/dcos\Theta$ in the
region $\zeta^{+} > \tilde{\zeta^{+}}$ ) from the Table 1 on
$(A_{P}*A_{T})^{1/2}$ is presented. One can see, that $T$ decreases
linearly with the increasing of $(A_{P}*A_{T})^{1/2}$ i.e with the increasing
number of participating nucleons. This decrease of the temparature with
increasing atomic number seems to be related to the fact that average
multiplicity of secondaries increases with increasing atomic number,
so the average
energy per secondary decreases.
\par
It is interesting to compare the temperatures extracted by means of our
analysis with those extracted in the GSI experiments (FOPI, KAON and TAPS-
Collaborations, see, e.g. [22,23]). The numerical values of the parameter
$T$ for pions in Au-Au collisions at 1 AGeV and our values for heaviest
colliding pair are close to each other. But it should be noted that the
extraction procedures are different. It seems interesting in this connection
to perform the light front analysis of the GSI data.

\  \par
\begin{center}
\bf { 5.  CONCLUSIONS }
\end{center}
\  \par
\par
The light front
analysis of $\pi^{-}$ -- mesons from $A_{P}$ (He, C, Mg, O) + $A_{T}$
(Li, C, Ne, Mg, Cu, Pb)
collisions has been carried out.
In some region of phase space of $\pi^{-}$
mesons
($\zeta^{+} >\tilde{\zeta^{+}}$)
the thermal equilibration seems to be reached,
 which is characterized by the temperature $T$.
The variables used can serve as a possible convenient tool to study
hadro-production processes in hadron-hadron,
nucleus-nucleus  and $e^{+}~e^{-}$ -- interactions.
\par
A remark on the nature of maxima in $\zeta^{\pm}$ -distributions is in order.
Recently ALEPH Collaboration observed the maxima in the $\xi$ - distributions
($\xi=-ln~ p/p_{max}$) [24] of secondary hadrons in $e^{+}~e^{-}$
collisions, which coincide to high precision with the predictions of
perturbative QCD (see., e.g. [25,26]). The accuracy of coincidence increases
when the next to leading order corrections are taken into account. So the
shapes of $\xi$ - distributions are related to the details of the underlying
dynamics. Similarly, it seems that the maxima in $\zeta^{\pm}$ -distributions
reflect the dynamics of the processes considered.
\   \par
\   \par
\   \par
\   \par
\   \par
ACKNOWLEDGEMENTS
\   \par
\   \par
\par
The authors express their deep gratitude to J.-P.Alard
N.Amaglobeli, Sh.Esakia, D.Ferenc, A.Golokhvastov, S.Khorozov, G.Kuratashvili, J.Lukstins,
J.-F.Mathiot, Z.Mente\\shashvili, G.Paic
for interesting discussions.
One of the authors (V.G.) would like to thank
A.De Rujula and G.Veneziano and CERN Theory Division for the warm hospitality.
\pagebreak
%\par
\   \par
\   \par
\   \par

\newpage
\begin{center}
\bf{FIGURE CAPTIONS}
\end{center}
\  \par
{\bf{Fig.1}}
Experimental set-up. The trigger and the trigger distances are not to
scale. \par
\  \par
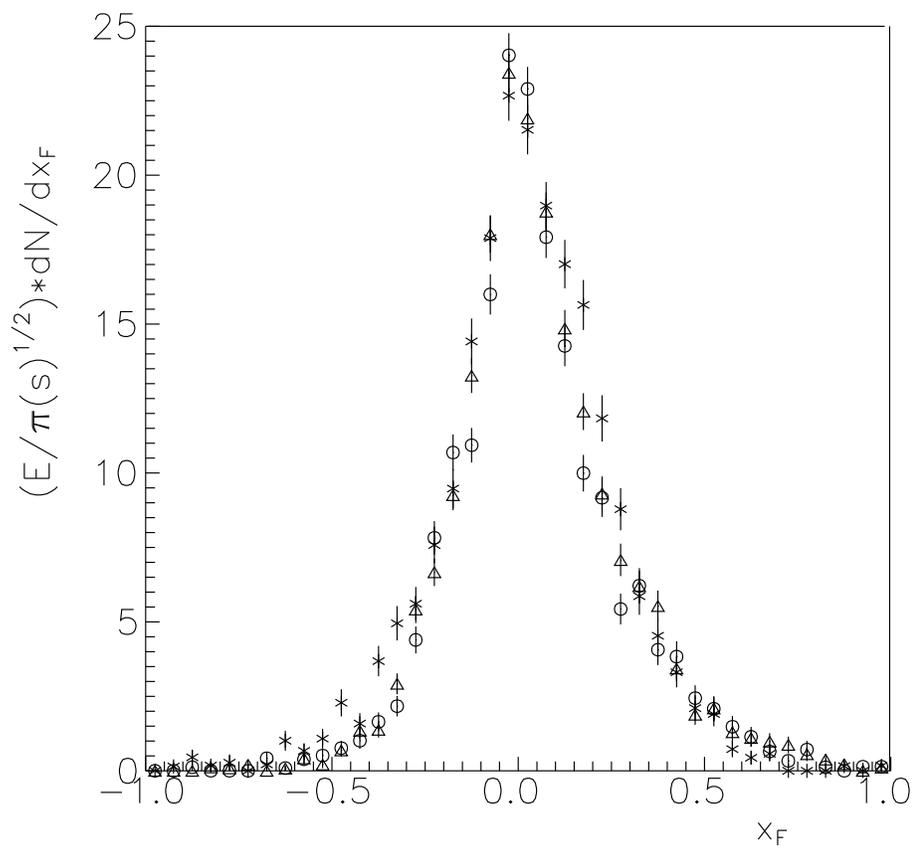
{\bf{Fig.2}}
The  $x_{F}$ -- distribution of  $\pi^{-}$ mesons from $\ast$ -- He(Li,C),
$\triangle$ -- C-Cu,  $\circ$   --  O-Pb
interactions.
\par
\   \par
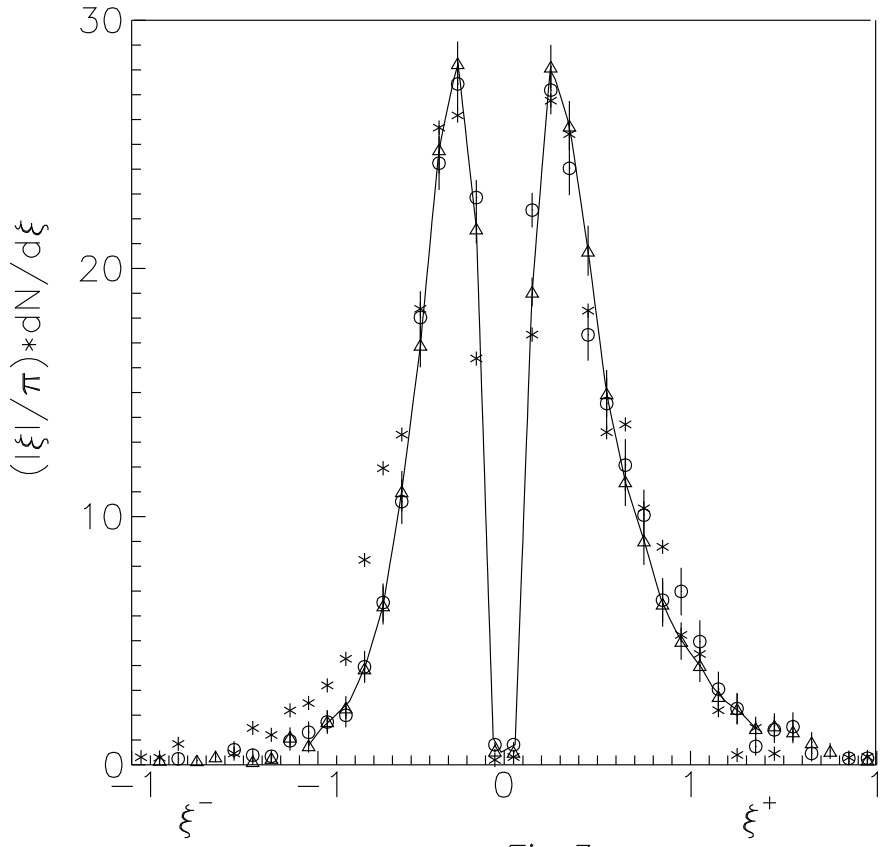
{\bf{Fig.3}}
The  $\xi^{\pm}$ -- distribution of  $\pi^{-}$ mesons
from $\ast$ -- He(Li,C),
$\triangle$ -- C-Cu,  $\circ$   --  O-Pb
interactions.
The curve is a result of polynomial approximation
of the experimental data.
\par
\   \par
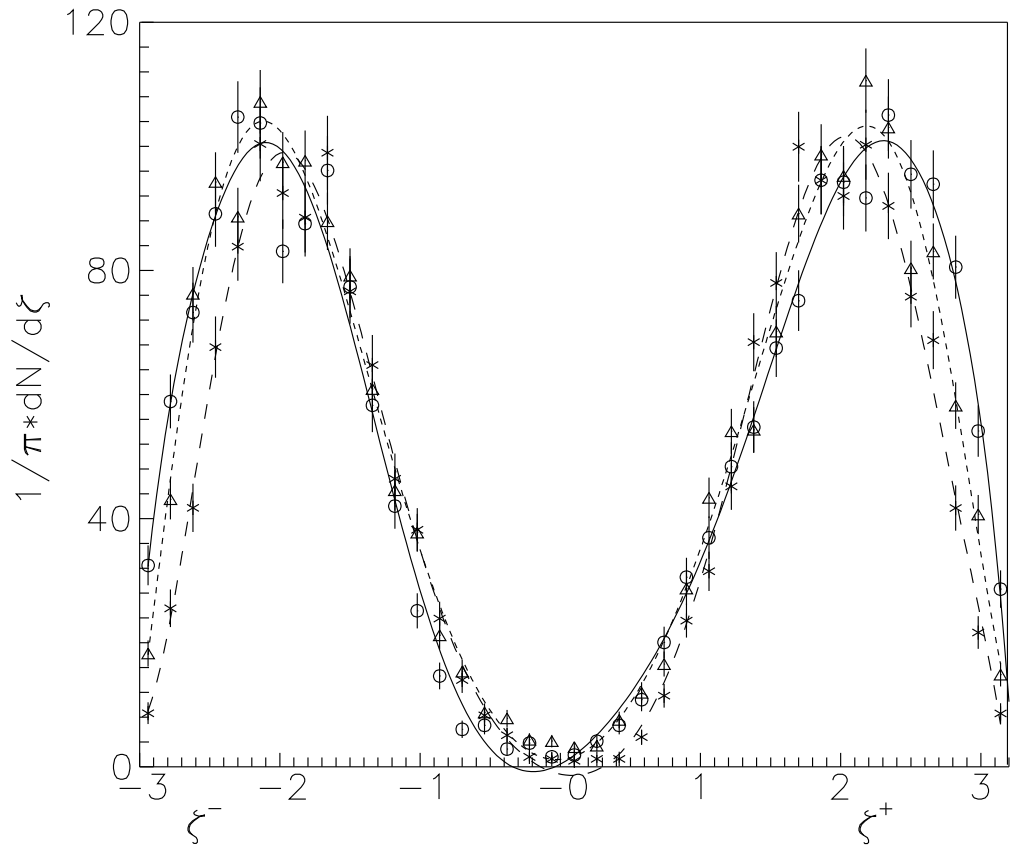
{\bf{Fig.4}}
The  $ \zeta^{\pm} $ -- distribution of  $\pi^{-}$ mesons from  $\ast$ -- He(Li,C),
$\triangle$ -- C-Cu,  $\circ$   --  O-Pb
interactions. The curves are the result of polynomial approximation.
\par
\  \par
{\bf{Fig.5.a}}
The $ p_{T}^{2} $ distribution of  $\pi^{-}$ mesons
from Mg-Mg interactions.
$\circ$   -- experimental data
for $\zeta^{+} > \tilde{\zeta^{+}}$   ($\tilde{\zeta^{+}}$=2.0),
$\bigtriangleup$ -- experimental data for $\zeta^{+} < \tilde{\zeta^{+}}$.
 The solid line - fit of the
data in the region $\zeta^{+} > \tilde{\zeta^{+}}$
by the eq.12 using the Boltzmann distribution
; the dashed line - fit of the  data in the region
$\zeta^{+} < \tilde{\zeta^{+}}$
by the eq.15.
\par
\   \par
{\bf{Fig.5.b}}
The $ p_{T}^{2} $ distribution of  $\pi^{-}$ mesons
from $\ast$ -- He(Li,C), $\triangle$ -- C-Cu,  $\circ$   --  O-Pb
interactions
for $\zeta^{+} > \tilde{\zeta^{+}}$   ($\tilde{\zeta^{+}}$=2.0).
 The solid line - fit of the
data in the region $\zeta^{+} > \tilde{\zeta^{+}}$
by the eq.12 using the Boltzmann distribution.
\par
\   \par
{\bf{Fig.5.c}}
The $ p_{T}^{2} $ distribution of  $\pi^{-}$ mesons
from $\ast$ -- He(Li,C), $\triangle$ -- C-Cu,  $\circ$   --  O-Pb
interactions for $\zeta^{+} < \tilde{\zeta^{+}}$.
The solid line - fit of the  data in the region
$\zeta^{+} < \tilde{\zeta^{+}}$
by the eq.15.
\par
\   \par
{\bf{Fig.6.a}}
The  $ cos\Theta $ distribution of  $\pi^{-}$ mesons
from Mg-Mg interactions.
$\circ$   -- experimental data
for $\zeta^{+} > \tilde{\zeta^{+}}$   ($\tilde{\zeta^{+}}$=2.0),
$\bigtriangleup$ -- experimental data for $\zeta^{+} < \tilde{\zeta^{+}}$.
 The solid line - fit of the
 data in the region $\zeta^{+} > \tilde{\zeta^{+}}$ by the eq.13
using the Boltzmann distribution;
the dashed line - fit of the
data in the region $\zeta^{+} < \tilde{\zeta^{+}}$
by the polynomial of the 6-th degree.
\par
\   \par
{\bf{Fig.6.b}}
The  $ cos\Theta $ distribution of  $\pi^{-}$ mesons
from $\ast$ -- He(Li,C), $\triangle$ -- C-Cu,  $\circ$   --  O-Pb
interactions
for $\zeta^{+} > \tilde{\zeta^{+}}$.
 The solid line - fit of the
 data in the region $\zeta^{+} > \tilde{\zeta^{+}}$ by the eq.13
using the Boltzmann distribution.
\par
\   \par
{\bf{Fig.6.c}}
The  $ cos\Theta $ distribution of  $\pi^{-}$ mesons
from $\ast$ -- He(Li,C), $\triangle$ -- C-Cu,  $\circ$   --  O-Pb
interactions for $\zeta^{+} < \tilde{\zeta^{+}}$.
 The solid line  - fit of the
data in the region $\zeta^{+} < \tilde{\zeta^{+}}$
by the polynomial of the 6-th degree.
\par
\   \par
{\bf{Fig.7.a}}
The  $ (1/\pi) \cdot dN/d\zeta^{+}  $ distribution of  $\pi^{-}$ mesons
from Mg-Mg interactions.
$\circ$ -- experimental data, the solid line -- fit of the data in the region
$\zeta^{+} > \tilde{\zeta^{+}}$
by the eq.11 using the Boltzmann distribution; the dashed line --
fit of the data in the region $\zeta^{+} < \tilde{\zeta^{+}}$  by the formula
$(1 - e^{-\vert \zeta^{+}\vert})^{n}$.
\par
\   \par
{\bf{Fig.7.b}}
The  $ (1/\pi) \cdot dN/d\zeta^{+}  $ distribution of  $\pi^{-}$ mesons
from C-Cu interactions.
$\circ$ -- experimental data, the solid line -- fit of the data in the region
$\zeta^{+} > \tilde{\zeta^{+}}$
by the eq.11 using the Boltzmann distribution; the dashed line --
fit of the data in the region $\zeta^{+} < \tilde{\zeta^{+}}$  by the formula
$(1 - e^{-\vert \zeta^{+}\vert})^{n}$.
\par
\   \par
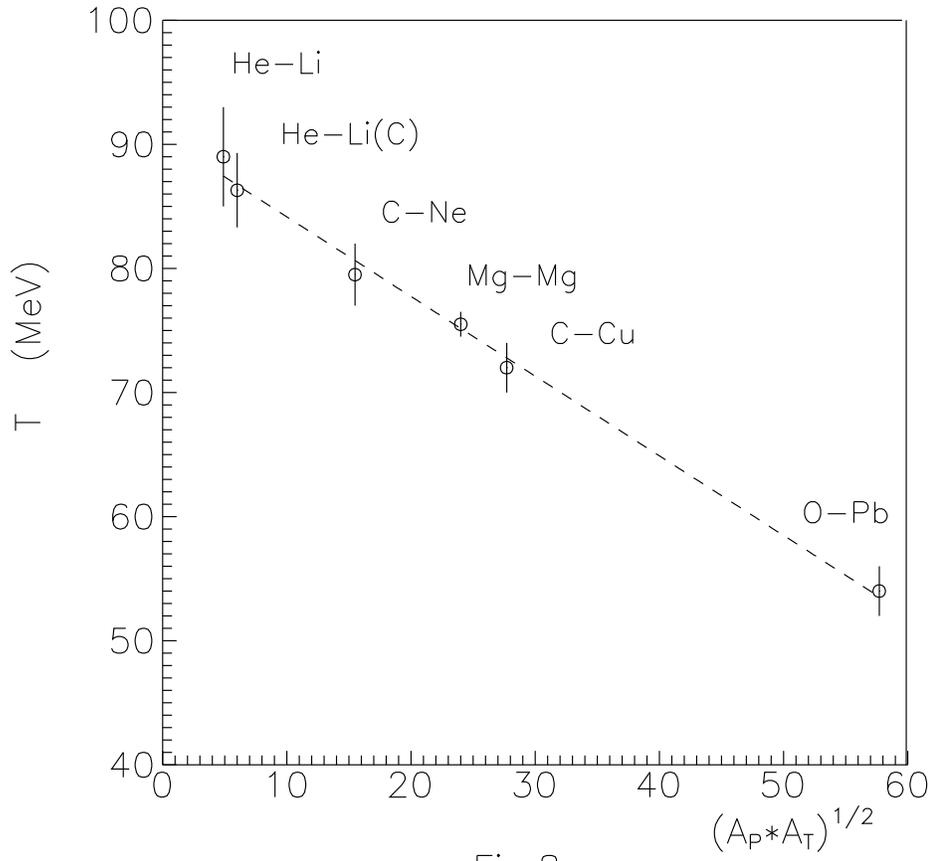
{\bf{Fig.8}}
 {The dependence of the  parameter $T$ on $(A_{P}*A_{T})^{1/2}$
for He-Li, He(Li,C), C-Ne, Mg-Mg, C-Cu and O-Pb. The dashed line is a
result of linear approximation.}
%\par
\   \par
\   \par
\begin{center}
\bf{TABLE CAPTIONS}
\end{center}
\   \par
{\bf{Table 1.}}
The
results of the fit of the distributions of $\pi^{-}$ --
mesons in the region  $\zeta^{+} >\tilde{\zeta^{+}}$.\par
\  \par
\  \par
{\bf{Table 2.}} The results of the fit of the  distributions of $\pi^{-}$
mesons in the region  $\zeta^{+} < \tilde{\zeta^{+}}$.
\newpage
{\bf{Table 1.}} Number of events, trigger, $\tilde{\zeta^{+}}$ and the
results of the fit of the experimental distributions of $\pi^{-}$ --
mesons in the region  $\zeta^{+} >\tilde{\zeta^{+}}$.\\
\  \par
\  \par
\  \par
\begin{tabular}{|c|c|c|c|c|c|c|c|c|}    \hline
%&  & &  & & & & &     \\
\multicolumn{1}{|c|}{}& \multicolumn{1}{c|}{}& \multicolumn{1}{c|}{}&
\multicolumn{6}{c|}{}\\
\multicolumn{1}{|c|}{$A_{p} - A_{T} $}
&\multicolumn{1}{c|}{$ Number $}&  \multicolumn{1}{c|}{}&
\multicolumn{6}{c|}{{$T~(MeV)$}}\\
\cline{4-9}
\multicolumn{1}{|c|}{}& \multicolumn{1}{c|}{of}& \multicolumn{1}{c|}{}&
\multicolumn{2}{c|}{}&\multicolumn{2}{c|}{}&\multicolumn{2}{c|}{}\\
\multicolumn{1}{|c|}{$T(\Theta_{ch},\Theta_{n}$)}&
\multicolumn{1}{c|}{ $ events$}&
\multicolumn{1}{c|}{$\tilde{\zeta ^{+}}$}&
\multicolumn{2}{c|}{$(1/\pi)\cdot dN/d\zeta ^{+}$}&
\multicolumn{2}{c|}{$dN/dp_{T}^{2}$}&
\multicolumn{2}{c|}{$dN/dcos\Theta $}\\
\multicolumn{1}{|c|}{}& \multicolumn{1}{c|}{}& \multicolumn{1}{c|}{}&
\multicolumn{2}{c|}{}&\multicolumn{2}{c|}{}&\multicolumn{2}{c|}{}\\
\cline{4-9}
\multicolumn{1}{|c|}{}&
\multicolumn{1}{c|}{}&
\multicolumn{1}{c|}{}&
\multicolumn{1}{c|}{$Boltz$}&
\multicolumn{1}{c|}{$B-E$}&
\multicolumn{1}{c|}{$Boltz$}&
\multicolumn{1}{c|}{$B-E$}&
\multicolumn{1}{c|}{$Boltz$}&
\multicolumn{1}{c|}{$B-E$}\\
\hline
& &&&&&&&\\
 $ He-Li $ &  4020 & 2.0 &  88 $\pm$ 6 &  88 $\pm$ 6
  & 85 $\pm$5 & 92 $\pm$7
  & 94 $\pm$ 12 & 102 $\pm$ 7 \\
$T(0,0)$&&&&&&&&\\
\hline
  $He(Li,C)$  & 6147 & 2.0 &  92 $\pm$ 5 &  92 $\pm$ 5
  & 84 $\pm$ 4 & 90 $\pm$ 5
   & 84 $\pm$ 9 & 90 $\pm$ 11 \\
$T(0,0)$&&&&&&&&\\
\hline
 $ C-Ne $ &  902 & 2.1 &  82 $\pm$ 7 & 82 $\pm$ 7
  & 75 $\pm$ 5 &  80 $\pm$ 6
   & 90 $\pm$ 14 & 94 $\pm$ 19 \\
$T(2,0)$&&&&&&&&\\
\hline
  Mg-Mg &  6239 & 2.0 &  75 $\pm$ 3 & 75 $\pm$ 4
& 76 $\pm$ 2 &  76 $\pm$ 3
   & 75 $\pm$ 3 & 78 $\pm$ 4\\
$T(2,2)$&&&&&&&&\\
\hline
 $ C-Cu $&  1203 & 2.0 & 73 $\pm$ 3 & 73 $\pm$ 3
  &  71 $\pm$ 2 &  74 $\pm$ 3
   & 73 $\pm$ 6 & 76 $\pm$ 7\\
$T(3,3)$&&&&&&&&\\
\hline
 $ O-Pb $&  732 & 2.0 & 55 $\pm$ 2 & 54 $\pm$ 1
  & 53 $\pm$ 2 &  53 $\pm$ 1
   & 54 $\pm$ 5 & 63 $\pm$ 4\\
$T(2,0)$&&&&&&&&\\
\hline
\end{tabular}
\newpage
{\bf{Table 2.}} The results of the fit of the experimental distributions of $\pi^{-}$
mesons in the region  $\zeta^{+} < \tilde{\zeta^{+}}$ .
\   \par
\   \par
\  \par
\  \par
\begin{tabular}{|c|c|c|c|c|}    \hline
\multicolumn{1}{|c|}{}&
\multicolumn{3}{|c|}{}&
\multicolumn{1}{c|}{}\\
\multicolumn{1}{|c|}{}&
\multicolumn{3}{|c|}{{$dN/dp_{T}^{2}$}}&
\multicolumn{1}{c|}{{$1/\pi \ast dN/d\zeta ^{+}$ }}\\
\multicolumn{1}{|c|}{{$A_{p}-A_{T}$}}&
\multicolumn{3}{|c|}{}&
\multicolumn{1}{c|}{}\\
\cline{2-5}
%&&&\\
   &$  \alpha $ & $\beta_{1}$ &$ \beta_{2}$  & $n$ \\
 &   &$ (GeV/c)^{-2}$ & $(GeV/c)^{-2}$ & \\
\hline
&&&&\\
 $ He-Li$ &  0.51$\pm$ 0.34 & 17.3 $\pm$ 1.9 & 7.5 $\pm$ 1.7 &3.79 $\pm$ 0.20  \\
&&&&\\
\hline
&&&&\\
 $He(Li,C)$ & 0.92 $\pm$ 0.15 & 11.1 $\pm$ 1.6 & 4.5 $\pm$ 1.3 &3.61 $\pm$ 0.15  \\
&&&&\\
\hline
&&&&\\
  $C-Ne$ &  0.46 $\pm$ 0.25 & 19.7 $\pm$ 1.8 & 8.4 $\pm$ 1.3 &2.04 $\pm$ 0.11 \\
&&&&\\
\hline
&&&&\\
 $Mg-Mg$ &  0.85 $\pm$ 0.03 & 12.0 $\pm$ 0.4 & 4.8 $\pm$ 0.3 &4.30 $\pm$ 0.06 \\
&&&&\\
\hline
&&&&\\
  $C-Cu$  & 0.86 $\pm$ 0.16 & 12.5 $\pm$ 2.1 & 4.9 $\pm$ 1.2 & 2.96 $\pm$ 0.11\\
&&&&\\
\hline
&&&&\\
  $O-Pb$ & 0.71 $\pm$ 0.30 & 12.5 $\pm$ 2.1 & 8.1 $\pm$ 0.4 &  2.59 $\pm$ 0.10 \\
&&&&\\
\hline
\end{tabular}
\end{document}